\documentclass[aps,pra,showpacs,notitlepage]{revtex4-1}

\usepackage{graphicx}

\begin{document}

\title{Quantum effects in the interference of photon number states}

\author{Holger F. Hofmann}
 \email{hofmann@hiroshima-u.ac.jp} 
 
\author{Keito Hibino}%

\author{Kazuya Fujiwara}%

\author{Jun-Yi Wu}%

\affiliation{
 Graduate School of Advanced Sciences of Matter, Hiroshima University
 1-3-1 Kagamiyama, Higashi-Hiroshima, 739-8530, Japan}

\begin{abstract}
Multi-photon interference results in modulations of output probabilities with phase shift periods that are much shorter than $2 \pi$. Here, we investigate the physics behind these statistical patterns in the case of well-defined photon numbers in the input and output modes of a two-path interferometer. We show that the periodicity of the multi-photon interference is related to the weak value of the unobserved intensity difference between the two arms of the interferometer. This means that the operator relations between the photon number differences in input, path, and output can be used to determine the periodicity of the experimentally observed quantum interference, establishing an important link between the classical causality of random phase interference and quantum effects that depend on the superposition of classically distinct possibilities.  
\end{abstract}


\maketitle

\section{Introduction}
One of the most basic operations of linear optics is the interference between two optical modes in a two-path interferometer. In quantum optics, this scenario has been widely studied in the context of quantum metrology, where it serves to illustrate the role of non-classical correlations in the phase sensitivity of multi-photon interference \cite{Hol93,Fiu02,Ste04,Uys07,Hof07,Dow08,Pez08,Hof09,Bir12,Roz14,Sah15,Pez15}. Unfortunately, these arguments about phase sensitivity often leave out the details of the physics, focusing merely on the perceived figures of merit and not on the mechanisms by which multi-photon interference patterns emerge. Recent experiments on multi-photon interference are providing a much more detailed picture of non-classical effects in two-mode interferences \cite{Mit04,Nag07,Sun08,Afe10,Isr12,Coo13,Xia13}, and it might be time to ask what these experimental observation of multi-photon fringes can tell us about the nature and origin of quantum interference. 

To properly address this question, it is necessary to distinguish carefully between classical wave interference and quantum interference. In typical experiments, multi-photon fringes are detected by measuring the phase dependence of count rates for a fixed combination of input and output conditions. Quantum interference thus appears as a modulation in the phase dependence of a single multi-photon probability. This is quite different from classical interference, where the output intensity is a deterministic function of the phase shift and probabilities appear only as a technical noise background. In quantum optics, classical interference is usually associated with the Poissonian photon counting statistics of coherent states, which describes a highly localized increase of detection probability for the classically expected intensity distribution at a specific phase shift in the interferometer. In the limit of high photon number, we can distinguish classical interference and multi-photon quantum interference qualitatively, since only the latter effect is associated with multiple interference fringes in the phase dependence of the output probabilities, and these interference fringes have distinct periodicities that are much shorter than $2 \pi$. It is therefore a non-trivial challenge to explain the relation between the modulations of probability seen in multi-photon interference and the statistics of classical interference effects observed in the phase dependence of output intensities. 

In the present paper, we address this problem by examining the role of the operators representing the coherence of the two field modes. These operators have a clear classical meaning, describing the intensity differences between any two modes as a component of a three dimensional vector in close analogy to the algebra of quantized spins. We point out that the phase dependence of the quantum state components representing a specific measurement outcome can be described in terms of weak values of the intensity difference between the two arms of the interferometer. These weak values can be written as a function of the eigenvalues of the initial and the final state using relations that correspond to the classical causality of two-path interference. For the case of photon number states in both the input and the output, we can then derive a differential equation describing the phase dependence of a single interference fringe, where the main contribution to the phase dependence originates from the rapid oscillation of the fringes, while the slowly varying envelope describes the statistics expected for a classical interference of two fields when the optical phases of the input fields are completely random. It is possible to identify the periodicity of the multi-photon fringes that describe the quantum mechanical modulation of output probability with the intensity difference between the paths obtained from the weak value of the squared operator representing this intensity difference. The periodicity of multi-photon interference fringes can thus be explained in terms of a classical estimate of the intensity difference between the paths obtained from the experimentally controlled input and output conditions. 

Our results show that multi-photon interference can be understood in terms of the classical relations between the intensities and coherences of the two modes. Specifically, the quantum mechanical phase of the multi-photon interference fringes is given by a classical action that relates the optical phase shift to its generator, the intensity difference between the paths. The non-classical effects of multi-photon interference can thus be traced to the fundamental role of the action in quantum physics \cite{Hof14,Hof16}. It may be worth noting that the direct identification of the action of phase shifts also provides a more efficient approach to the quantization of fields that starts from the macroscopically observable physics and hence avoids many of the ambiguities associated with the mathematical concepts of superpositions and state vectors. Multi-photon interference may thus help us bridge the gap between quantum mechanical concepts and classical intuition in a new and unexpected manner.  

The paper is organized as follows: In Sec. \ref{sec:phase}, we review the operator algebra of two-path interference and its relation to the interference fringes observed in multi-photon experiments. In Sec. \ref{sec:wv}, it is shown that weak values can be used to express the phase dependence of a single multi-photon interference fringe. Based on this observation, we derive a differential equation for the interference fringes obtained with well-defined photon numbers in the input and the output. In Sec. \ref{sec:separate}, the differential equation derived in Sec. \ref{sec:wv} is solved approximately by separating the interference fringe into an interference term described by a phase dependent action $S$ and an envelope function $A$. It is shown that the action evolves according to a classical Hamilton-Jacobi equation, while the envelope describes the statistics of classical random phase interference. In Sec. \ref{sec:examples} we apply the theory to a number of characteristic cases, comparing the approximate results to exact results obtained for eight and sixteen photons. The results show that the approximation correctly describes the main features of multi-photon interference, especially with regard to the separation of quantum interference effects and classical random phase field interference. In Sec. \ref{sec:action} we identify the necessary conditions for the identification of multi-photon interference fringes with a classical action function. It is shown that weak values can be used to identify the action of multi-photon phase interferences for a wide variety of possible input states. In Sec. \ref{sec:NOON} we discuss the relation between the general analysis of multi-photon interference using the action and the definition of multi-photon interference based on superpositions of photon number eigenstates in the interferometer paths associated with the well known NOON states. It is pointed out that the NOON states represent a special case of the general action-based theory. Sec. \ref{sec:concl} summarizes the results and concludes the paper.

\section{Phase shifts in two-path interferometers}
\label{sec:phase}

The physics of two optical modes can be described in terms of the field operators $\hat{a}$ and $\hat{b}$ that describe the complex field amplitudes of the two modes. Due to their mathematical effects on photon number states, these field operators are commonly known as annihilation operators, although it should be kept in mind that this mathematical effect is not related to the physical properties described by the operators in any obvious or intuitive manner. The proper connection between two mode coherences and photon number is obtained by considering the second order products of field amplitudes,
\begin{eqnarray}
\hat{J}_1 &=&\frac{\hbar}{2}\left(\hat{a}^\dagger \hat{b} + \hat{b}^\dagger\hat{a}\right)
\nonumber \\
\hat{J}_2 &=& -i \frac{\hbar}{2}\left(\hat{a}^\dagger \hat{b} - \hat{b}^\dagger\hat{a}\right)
\nonumber \\
\hat{J}_3 &=&\frac{\hbar}{2}\left(\hat{a}^\dagger \hat{a} - \hat{b}^\dagger\hat{b}\right).
\end{eqnarray}
Each component of this three dimensional vector represents an intensity difference between two orthogonal modes in units of $\hbar/2$ per photon. The motivation for this choice of units is the representation of phase shifts between the modes $\hat{a}$ and $\hat{b}$, which is generated by the component $\hat{J}_3$ in the same way that a Hamiltonian generates the time evolution. Specifically, the unitary transformation of a phase shift is given by
\begin{equation}
\label{eq:unitary}
\hat{U}(\phi) = \exp\left(-i\frac{1}{\hbar} \hat{J}_3 \phi \right)
\end{equation}
and the effect of a phase shift on an arbitrary state $\mid \psi \rangle$ can be described by 
\begin{equation}
\label{eq:Schroed}
\frac{\partial}{\partial \phi} \mid \psi \rangle = -i \frac{1}{\hbar} \hat{J}_3 \mid \psi \rangle.
\end{equation}
Here, the product $\hat{J}_3 \phi$ in Eq. (\ref{eq:unitary}) represents the action of a phase shift, just as the action of a time evolution is given by the energy-time product $\hat{H} t$. 

In a two path interferometer, $\hat{J}_3$ describes the intensity difference between the two paths in the interferometer. The intensity difference between the input modes is usually given by $\hat{J}_1$. The remaining component $\hat{J}_2$ describes the phase coherence between the input fields that results in interferences when a phase shift of $\phi$ is applied. For a phase shift of $\phi$, the intensity difference observed in the output can be given by
\begin{eqnarray}
\label{eq:Heis}
\hat{J}_\phi &=& \hat{U}^\dagger(\phi) \hat{J}_1 \hat{U}(\phi)
\nonumber \\
 &=& \cos(\phi) \hat{J}_1 - \sin(\phi) \hat{J}_2.
\end{eqnarray}
In the following, we will consider input states $\mid \psi (\phi=0) \rangle$ that are eigenstates of the input intensity difference $\hat{J}_1$, and measurements of the output intensity difference $\hat{J}_\phi$. Specific measurement results can be given by an integer or half-integer value of $m$ representing one half of the photon number difference between the output ports. Since the total photon number $N$ is conserved, $m$ represents a photon number state of the two output ports with output photon numbers of $N/2+m$ and $N/2-m$. The quantum states $\{\mid m \rangle \}$ representing these measurement results are eigenstates of $\hat{J}_\phi$ with eigenvalues of $\hbar m$. We are therefore interested in the phase dependent values of the output probabilities given by 
\begin{equation}
P(m;\phi) = |\langle m \mid \hat{U}(\phi) \mid \psi (0) \rangle|^2,
\end{equation}
where the unitary transformation is used to relate the state $\mid \psi(\phi)\rangle$ to the input state $\mid \psi(0)\rangle$. Experimentally, these probabilities are obtained as multi-photon coincidence rates for the detection of $N/2-m$ photons in one output port and $N/2+m$ photons in the other. Quantum interference is observed as a phase dependent modulation of the probability, with a periodicity that can be as short as $2 \pi/N$ for $N$ photons. We will now analyze the phase dependence of $\langle m \mid \psi (\phi) \rangle$ to identify the origin of these quantum interference fringes in multi-photon interference.  

\section{Derivation of multi-photon interference fringes using weak values}
\label{sec:wv}

We can apply the transformation of quantum states given by Eq.(\ref{eq:Schroed}) to describe the phase evolution of the probability amplitudes $\langle m \mid \psi (\phi) \rangle$ associated with specific measurement outcomes $m$. In its most conventional form, the resulting differential equation is given by
\begin{equation}
\frac{\partial}{\partial \phi} \langle m \mid \psi(\phi) \rangle = -i \frac{1}{\hbar} \langle m \mid \hat{J}_3 \mid \psi(\phi) \rangle.
\end{equation}
In the textbook approach to quantum dynamics, the operator $\hat{J}_3$ is usually expanded into its matrix representation in the measurement basis $\{\mid m \rangle \}$. However, this may not be the most meaningful analysis of the role of the generator $\hat{J}_3$ in the evolution of the input-output relation $\langle m \mid \psi \rangle $. A closer correspondence to classical dynamics can be maintained by identifying the contribution of $\hat{J}_3$ with its weak value,
\begin{equation}
\label{eq:action}
\frac{\partial}{\partial \phi} \langle m \mid \psi(\phi) \rangle = -i \frac{1}{\hbar} \frac{\langle m \mid \hat{J}_3 \mid \psi(\phi) \rangle}{\langle m \mid \psi(\phi) \rangle} \langle m \mid \psi(\phi) \rangle.
\end{equation}
It is interesting to note that the weak value can be used to express the effects of the operator $\hat{J}_3$ on the phase evolution of the state component $\langle m \mid \psi(\phi) \rangle$. The advantage of this approach is that it is often possible to find the weak value without having to solve the complete dynamics in the Schr\"odinger picture. Specifically, weak values can be determined by expressing the operator as a function of two operators, where the initial state is an eigenstate of the first and the final state is an eigenstate of the second operator. 

In the present case, $\mid \psi (\phi=0) \rangle$ is an eigenstate of $\hat{J}_1$ and $\mid m \rangle$ is an eigenstate of $\hat{J}_\phi$, which can be related to the operators at $\phi=0$ using Eq.(\ref{eq:Heis}). Since the algebra of the $\hat{J}_i$ operators is the familiar algebra of spin operators for a total spin quantum number of $l=N/2$, it is possible to derive a relation between the phase shift generator $\hat{J}_3$ and the components $\hat{J}_1$ and $\hat{J}_\phi$ in the $\hat{J}_1$-$\hat{J}_2$ plane orthogonal to $\hat{J}_3$ by using the total length of the $J$-vector,
\begin{eqnarray}
\label{eq:Jquad}
\hat{J}_3^2 &=& \frac{\hbar^2}{4} N(N+2) - \hat{J}_1^2 - \hat{J}_2^2
\nonumber \\
&=& \frac{\hbar^2}{4} N(N+2) - \frac{1}{(\sin(\phi))^2}\left(\hat{J}_1^2 - \cos(\phi) (\hat{J}_1 \hat{J}_\phi + \hat{J}_\phi \hat{J}_1) + \hat{J}_\phi^2\right).
\end{eqnarray}
Since the initial and the final state are eigenstates of $J$-vector components orthogonal to $\hat{J}_3$, it is not possible to distinguish negative values of $\hat{J}_3$ from positive values of $\hat{J}_3$. As a result of this symmetry, the real part of the weak value of $\hat{J}_3$ is zero at all phases $\phi$, and the ratio between the amplitude $\langle m \mid \psi (\phi) \rangle$ and its phase derivative in Eq.(\ref{eq:action}) is always real. It is therefore possible to express $\langle m \mid \psi (\phi) \rangle$ by real numbers for all values of $\phi$. 

To make optimal use of the relation in Eq.(\ref{eq:Jquad}), we now consider the second derivative of the phase dependence,
\begin{eqnarray}
\label{eq:quadact}
\frac{\partial^2}{\partial\phi^2} \langle m \mid \psi(\phi) \rangle &=& - \frac{1}{\hbar^2} \langle m \mid \hat{J}^2_3 \mid \psi(\phi) \rangle
\nonumber \\
&=& - \frac{1}{\hbar^2} \frac{\langle m \mid \hat{J}^2_3 \mid \psi(\phi) \rangle}{\langle m \mid \psi(\phi) \rangle} \langle m \mid \psi(\phi) \rangle.
\end{eqnarray}
This derivative is described by the weak value of $\hat{J}_3^2$, and this weak value can be determined using the eigenvalues $m_\psi$ and $m$ for the input state and the measurement outcome, respectively. It is important to arrange the order of the operators so that the operator $\hat{J}_1$ is always to the right of the operator $\hat{J}_\phi$, and the necessary application of the commutation relations results in a contribution from the imaginary weak value of $\hat{J}_3$. The final relation between the weak values and the eigenvalues therefore reads
\begin{equation}
\frac{\langle m \mid \hat{J}^2_3 \mid \psi(\phi) \rangle}{\langle m \mid \psi(\phi) \rangle} 
+i \hbar \frac{\cos(\phi)}{\sin(\phi)} \frac{\langle m \mid \hat{J}_3 \mid \psi(\phi) \rangle}{\langle m \mid \psi(\phi) \rangle} 
= \frac{\hbar^2}{4} N(N+2) - \frac{\hbar^2}{(\sin(\phi))^2}\left(m_\psi^2 - 2 \cos(\phi) m_\psi m + m^2\right).
\end{equation}
Since both the weak value of $\hat{J}_3^2$ and the weak value of $\hat{J}_3$ appear in the phase derivatives of $\langle m \mid \psi(\phi) \rangle$, we can use this relation between the weak values and the eigenvalues to find a differential equation for the phase dependence of the probability amplitude $\langle m \mid \psi(\phi) \rangle$ that does not depend on the probability amplitudes of any other measurement outcomes $m$. According to Eqs.(\ref{eq:action}) and (\ref{eq:quadact}), this differential equation can be written as
\begin{equation}
\label{eq:fringes}
\frac{\partial^2}{\partial\phi^2} \langle m \mid \psi(\phi) \rangle
+ \frac{\cos(\phi)}{\sin(\phi)} \frac{\partial}{\partial\phi} \langle m \mid \psi(\phi) \rangle
= - \left(\frac{1}{4} N(N+2) - \frac{1}{(\sin(\phi))^2}\left(m_\psi^2 - 2 \cos(\phi) m_\psi m + m^2\right)\right) \langle m \mid \psi(\phi) \rangle.
\end{equation}
We have thus derived a general mathematical description of the multi-photon interference fringes observed with any photon number input. In the following, we will consider the physics described by this differential equation and identify the characteristic features of its solutions. 

\section{Separation of quantum effects and field statistics}
\label{sec:separate}

The description of the effects of phase shifts on multi-photon states given by Eq.(\ref{eq:fringes}) makes it possible to separate quantum effects from the classical limit by considering how the relation changes with total photon number. Classical effects should all scale with the total intensity, whereas the magnitude of quantum effects will always depend on absolute photon numbers. In the present case, such a separation of scales can be achieved by expressing the probability amplitudes $\langle m \mid \psi(\phi) \rangle$ by a product of a slowly varying envelope function $A(\phi)$ and a quantum interference effect given by an action $S(\phi)$ that describes the rapid modulation of probability associated with multi-photon interference,
\begin{equation}
\label{eq:separate}
\langle m \mid \psi(\phi) \rangle = 2 A(\phi) \cos\left(\frac{1}{\hbar} S(\phi)\right).
\end{equation}
The factor of two represents the interference between two classical solutions, as will be seen more clearly in the following discussion. Note that there is no approximation involved at this point, and the separation in Eq.(\ref{eq:separate}) can be used to obtain an exact solution of Eq.(\ref{eq:fringes}) for a specific photon number. However, our main concern is the comparison between the quantum effects that are expressed by the rapid oscillation of $\cos(S/\hbar)$ and the much slower variation of $A(\phi)$ that corresponds more closely to the classical statistics of random phase interference. In the following, we will therefore focus on approximate solutions, where the different scales of the phase dependence allow a complete separation between the dynamics of $S(\phi)$ and the dynamics of $A(\phi)$.
Specifically, the use of $\hbar$ in Eq.(\ref{eq:separate}) allows us to expand Eq.(\ref{eq:fringes}) in $\hbar$, where the classical limit emerges when action differences of $\hbar$ are not resolved. In the limit of sufficiently large photon numbers, we can therefore identify quantum effects as effects that depend on the precise ratio between the macroscopic action $S(\phi)$ and the fundamental constant $\hbar$. 

In Eq.(\ref{eq:fringes}), the derivatives result in contributions that depend explicitly on the fundamental constant $\hbar$. Due to the smallness of $\hbar$, we can expand the equation and neglect higher order contributions of $\hbar$ in favor of the lower order contributions. The leading terms of the expansion are proportional to $1/\hbar^2$, which means that the left hand side of Eq. (\ref{eq:fringes}) can be represented by the square of the derivative of $S(\phi)$,
\begin{equation}
\label{eq:second}
\frac{\partial^2}{\partial\phi^2} \langle m \mid \psi(\phi) \rangle \approx - \left(\frac{1}{\hbar} \frac{\partial}{\partial \phi} S(\phi) \right)^2 \langle m \mid \psi(\phi) \rangle
\end{equation}
Note that the largest contribution of the first derivative of $\langle m \mid \psi(\phi) \rangle$ in Eq.(\ref{eq:fringes}) is proportional to $1/\hbar$, so that the contribution of the first derivative to the left hand side of Eq.(\ref{eq:fringes}) can be neglected except at phases where the leading contribution of $\partial S/\partial \phi$ becomes very small. For non-vanishing values of $\partial S/\partial \phi$, the approximate solution of the equation is given by
\begin{equation}
\label{eq:Sapprox}
\frac{\partial}{\partial \phi} S(\phi) \approx - \sqrt{\frac{\hbar^2}{4} N(N+2) - \frac{\hbar^2}{(\sin(\phi))^2}\left(m_\psi^2 - 2 \cos(\phi) m_\psi m + m^2\right)},
\end{equation}
where the choice of sign is arbitrary since the action $S$ is defined by the phase of a cosine in Eq.(\ref{eq:separate}). We choose the negative sign because it corresponds to the conventional definition of action in classical Hamilton-Jacobi theory, as will be seen below. Importantly, the right hand side of this equation has a very intuitive physical meaning: it is the classical value of $\hat{J}_3$ obtained from the length of the $J$-vector and the values of $\hat{J}_1$ and $\hat{J}_\phi$ when the Heisenberg relation between the operators in Eq.(\ref{eq:Heis}) is converted to a classical relation between the eigenvalues by neglecting the non-commutativity of $\hat{J}_\phi$ and $\hat{J}_1$. We can define this classical approximation of $J_3$ as
\begin{equation}
\label{eq:J3}
J_3(\hbar m, \hbar m_\psi, \phi) = \sqrt{\hbar \frac{N}{2} \hbar\left(\frac{N}{2}+1\right) - \frac{1}{(\sin(\phi))^2}\left((\hbar m_\psi)^2 - 2 \cos(\phi) (\hbar m_\psi)(\hbar m) + (\hbar m)^2\right)},
\end{equation}
which describes the intensity difference between the arms of a classical two-path interferometer when an input intensity difference of $\hbar m_\psi$ and a phase shift of $\phi$ result in an output intensity difference of $\hbar m$. The approximate quantum mechanical solution to Eq. (\ref{eq:fringes}) given by Eq.(\ref{eq:Sapprox}) then corresponds to the classical relation between the generator $J_3$ and the action $S$ given by a version of the Hamilton-Jacobi equation,
\begin{equation}
\label{eq:HJE}
\frac{\partial}{\partial \phi} S(\hbar m, \hbar m_\psi, \phi)
= - J_3(\hbar m, \hbar m_\psi, \phi).
\end{equation}
The reason why this action describes a quantum effect is that it appears as the quantum phase of the multi-photon interference fringes described by Eq.(\ref{eq:separate}), where the classical action $S$ is converted into a quantum phase by dividing it by $\hbar$. Thus, the larger the classical action $S$ becomes, the faster the output probability oscillates, resulting in microscopic (and therefore highly phase sensitive) modulations of probability in the limit of macroscopic action. The approximations used above apply whenever the modulation dominates the phase dependence, which is the case whenever $J_3$ is sufficiently larger than $\hbar$. 

We can now turn to the approximate solution for the slowly varying envelope function $A(\phi)$. Since the quantum effects are for the most part described by the interference effects associated with the action in Eq.(\ref{eq:separate}), we expect the square of the envelope function $A(\phi)$ to represent the classical probability density of the multi-photon interference scenario. Specifically, the phase dependence of $A(\phi)$ can be determined by considering the second largest term in the expansion of Eq.(\ref{eq:fringes}) in $\hbar$, which is proportional to $1/\hbar$. The result can be written as
\begin{equation}
-\frac{2}{\hbar}\left(A(\phi) \frac{\partial^2}{\partial\phi^2} S(\phi) + 2  \left(\frac{\partial}{\partial\phi} A(\phi) \right)\left( \frac{\partial}{\partial\phi} S(\phi) \right) + \frac{\cos(\phi)}{\sin(\phi)} A(\phi) \frac{\partial}{\partial\phi} S(\phi)\right)\sin\left(\frac{S(\phi)}{\hbar}\right) = 0.
\end{equation}
The general solution of this equation does not depend on the specific value of $J_3$ and can be expressed by a relation between $A(\phi)$ and $S(\phi)$ given in more compact form as
\begin{equation}
\frac{\partial}{\partial\phi}\left(\sin(\phi) (A(\phi))^2 \frac{\partial}{\partial\phi} S(\phi) \right)=0.
\end{equation} 
Since the phase derivative of the action is given by (\ref{eq:HJE}), it is possible to express the squared envelope function $A^2$ as a function of $J_3$ and a normalization factor of $\rho_0$,
\begin{equation}
\label{eq:ergo}
A^2 (\hbar m, \hbar m_\psi, \phi) = \frac{\hbar \; \rho_0}{|\sin(\phi) J_3(\hbar m, \hbar m_\psi, \phi)|}.
\end{equation}
This relation corresponds to a classical density of output intensities, where the factor of $\hbar$ originates from the quantized distance between two eigenvalues of $\hat{J}_1$. Once the factor of $\hbar$ is removed, the probability density scales with total intensity, indicating that the envelope function $A$ represents statistics that are also observable in the classical limit. 

It is in fact possible to derive an expression for $\rho_0$ based on classical considerations. If we assume that the probability distribution originates from random phase interference between the two input intensities, we can explain the probability densities in terms of the phase difference $\theta$ between the input modes. Since the output value of $\hat{J}_\phi$ is a deterministic function of the fixed input value of $\hat{J}_1$ and the random phase $\theta$, the ratio of probability densities for intervals of $d \theta$ and intervals of $d J_\phi$ is given by
\begin{equation}
\left| \frac{d J_\phi}{d \theta} \right| = \left|\sin(\phi) J_3 \right|.
\end{equation}  
Thus the factor in the denominator of Eq.(\ref{eq:ergo}) effectively converts probability densities in the random phase $\theta$ into probability densities in the output observable $J_\phi$. For homogeneous phase distributions, the factor of $\rho_0$ in Eq.(\ref{eq:ergo}) is given by
\begin{equation}
\rho_0=\frac{d\rho}{d\theta} = \frac{1}{2 \pi}.
\end{equation}
Using these relations, it is possible to find a properly normalized solution for the envelope function $A(\phi)$.
In the light of this result, we can conclude that the slowly varying envelope function describes the classical statistics originating from the random optical phase of the photon number states in the input. Specifically, the phase dependence of $A^2$ given by Eq.(\ref{eq:ergo}) is independent of the absolute photon number if the photon number differences of $2m$ and $2 m_\psi$ are given by constant fractions of total photon number. In the limit of high $N$, this means that the shape of the envelope $A(\phi)$ does not change much if the photon number changes by about $\sqrt{N}$, corresponding to the Poissonian photon number fluctuations of coherent states. It is therefore also possible to observe $A(\phi)$ using random phase interferences between two coherent states of light, where the relative error in $m/N$ caused by photon number fluctuations drops to zero with $1/\sqrt{N}$. On the other hand, the multi-photon interference patterns described by $S(\phi)$ disappear in the statistics of coherent states because their periodicities are much shorter than the absolute shot noise value of $\sqrt{N}$ in the limit of high photon number. We can therefore confirm that $A(\phi)$ is classical in the sense that it can also be observed using input states that can be represented by mixtures of coherent states as described by a positive P-function, while the multi-photon fringes described by $S(\phi)$ are non-classical since they disappear when the input can be described as a mixture of coherent states.

In combination with the solution for the action $S$, we now have an approximate solution for $\langle m \mid \psi(\phi) \rangle$ that is valid whenever the values of $J_3$ are sufficiently large. For multi-photon states, this condition is usually satisfied if there is a classical solution that relates the input intensity to the output intensity for a phase shift of $\phi$. The condition only breaks down when the argument in the square root of Eq.(\ref{eq:J3}) is very small or negative. Note that Eq.(\ref{eq:fringes}) can also be solved in the regime of negative $J_3^2$ values, where the differential equation describes evanescent solutions characterized by an exponential suppression of the amplitude $\langle m \mid \psi(\phi) \rangle$ with increasing distance from the region of positive $J_3^2$. Mathematically, this situation is equivalent to the evanescent wavefunctions known from tunneling. In fact, a comparison of Eq.(\ref{eq:fringes}) with the time independent Schr\"odinger equation shows that the mathematics are quite similar, and that the method of approximation used above corresponds to the WKB-approximation in the region where the energy is larger than the potential energy. The regions where $J_3$ drops to zero and becomes imaginary can therefore be treated in close analogy to the WKB approximation for the tunneling problem, by allowing evanescent solutions where the cosine function in Eq.(\ref{eq:separate}) is replaced by a corresponding exponential function. The intermediate regime close to $J_3=0$ corresponds to the turning points of the WKB solution and can be approximated by Airy functions. In this manner, it is possible to obtain an approximate solution of Eq.(\ref{eq:fringes}) at all photon numbers. However, we are mostly concerned with the multi-photon interference fringes observed for $J_3 \gg 1$, so a detailed analysis of the evanescent parts of the phase dependence is not necessary in the present context.  

The main merit of the present analysis is that it allows us to separate the quantum interference effects described by the action $S$ from the effects of classical field statistics originating from the randomness of the optical phases in the input, which are represented by the envelope function $A$. In the following, we will apply this analysis to several representative examples of multi-photon interference fringes in order to illustrate how the approximations used here reproduce the characteristic features of multi-photon interference for different input-output relations. 

\section{Multi-photon fringes for different photon number distributions}
\label{sec:examples}

In the measurement scenarios we are considering, the input photon number distribution and the output photon number distribution are both fixed, and the phase dependence of the probability for this specific combination is investigated. For a given total photon number $N$, different measurement scenarios are defined by the combination of input photon number difference $2 m_\psi$ and output photon number difference $2 m$. To illustrate our method of analysis, it is best to start with a particularly simple class of scenarios, where the photon numbers in both input and output ports are equal, so that $m_\psi=m=0$. In this case, we find that the intensity difference between the two arms of the interferometer does not depend on phase and is simply given by the total length of the $J$-vector,
\begin{equation}
\label{eq:HB}
|J_3(0,0,\phi)| = \frac{\hbar}{2} \sqrt{N(N+2)} \approx \frac{\hbar}{2}(N+1).
\end{equation}
In terms of classical wave interference, it is easy to understand why equal intensities in both the input and the output require a maximal intensity difference between the paths. Since both $\hat{J}_1$ and $\hat{J}_\phi$ are zero, we can immediately conclude that $\hat{J}_2$ should also be zero, so that the $J$-vector must point in the direction of $\hat{J}_3$. It is also possible to argue that we do not observe any interferences between the two paths, and this is only possible when all of the light travels along only one of the two paths. As mentioned above, $J_3$ is actually derived from the weak value of $\hat{J}_3^2$, so the choice of sign in Eqs.(\ref{eq:J3}) and (\ref{eq:HJE}) is merely an arbitrary convention. In terms of the actual physics, the scenario is complete symmetric in the two paths, so it is impossible to distinguish positive and negative values of $J_3$. The quantum interferences described by $\cos(S/\hbar)$ in Eq.(\ref{eq:separate}) are a consequence of this path symmetry \cite{Hof09}. According to Eq.(\ref{eq:HJE}), the distance $\Delta \phi$ between two minima of the interference fringes ($\Delta S = \hbar \pi$) is given by
\begin{equation}
\Delta \phi = \frac{\hbar \pi}{|J_3|} = \frac{2 \pi}{(N+1)}.
\end{equation}
It is worth noting that the value of $|J_3|$ is larger than the maximal eigenvalue of $\hat{J}_3$, resulting in fringes that have a shorter period than fringes observed for superpositions of $m_3=N/2$ and $m_3=-N/2$, the states known as NOON states \cite{Dow08}. The observation of shorter fringes is somewhat surprising because firstly, the NOON states achieve the maximal possible phase sensitivity of $N$-photon states and secondly the fringes of photon number states can be written as a sum of fringes from superpositions of $m_3$ and $-m_3$, where the NOON state contribution is the contribution with the shortest fringe periodicity. However, the weak value of $\hat{J}_3^2$ is not limited by the available range of eigenvalues, resulting in a fringe periodicity that corresponds to an additional one photon intensity difference between the paths. Specifically, the classical relation between $J_3$ and the eigenvalues $\hbar m_\psi$ and $\hbar m$ does not include any uncertainties, and the total length of the $J$-vector is given by $N+1$ because the ``$+1$'' contribution represents the necessary quantum fluctuations of an eigenstate with maximal $\hat{J}_3$ eigenvalue. The shorter fringes therefore represent the absence of quantum noise in the deterministic relation between input photon numbers and output photon numbers. 
Importantly, the shorter fringes are only obtained for this specific combination of input and output, 
which means that this effect is compensated by measurement results with less phase sensitivity and does not result in an enhanced overall phase sensitivity \cite{Xia13,Hof12}. In fact, the experimental results reported in \cite{Xia13} confirm the observation of fringes shorter than the NOON state limit of $2 \pi/N$. It is interesting that the significance of this experimental result seems to have escaped notice, probably because there was no proper theoretical explanation available at the time. 

\begin{figure}[th]
\vspace{-1.5cm}
\begin{picture}(500,480)
\put(0,0){\makebox(480,480){
\scalebox{0.8}[0.8]{
\includegraphics{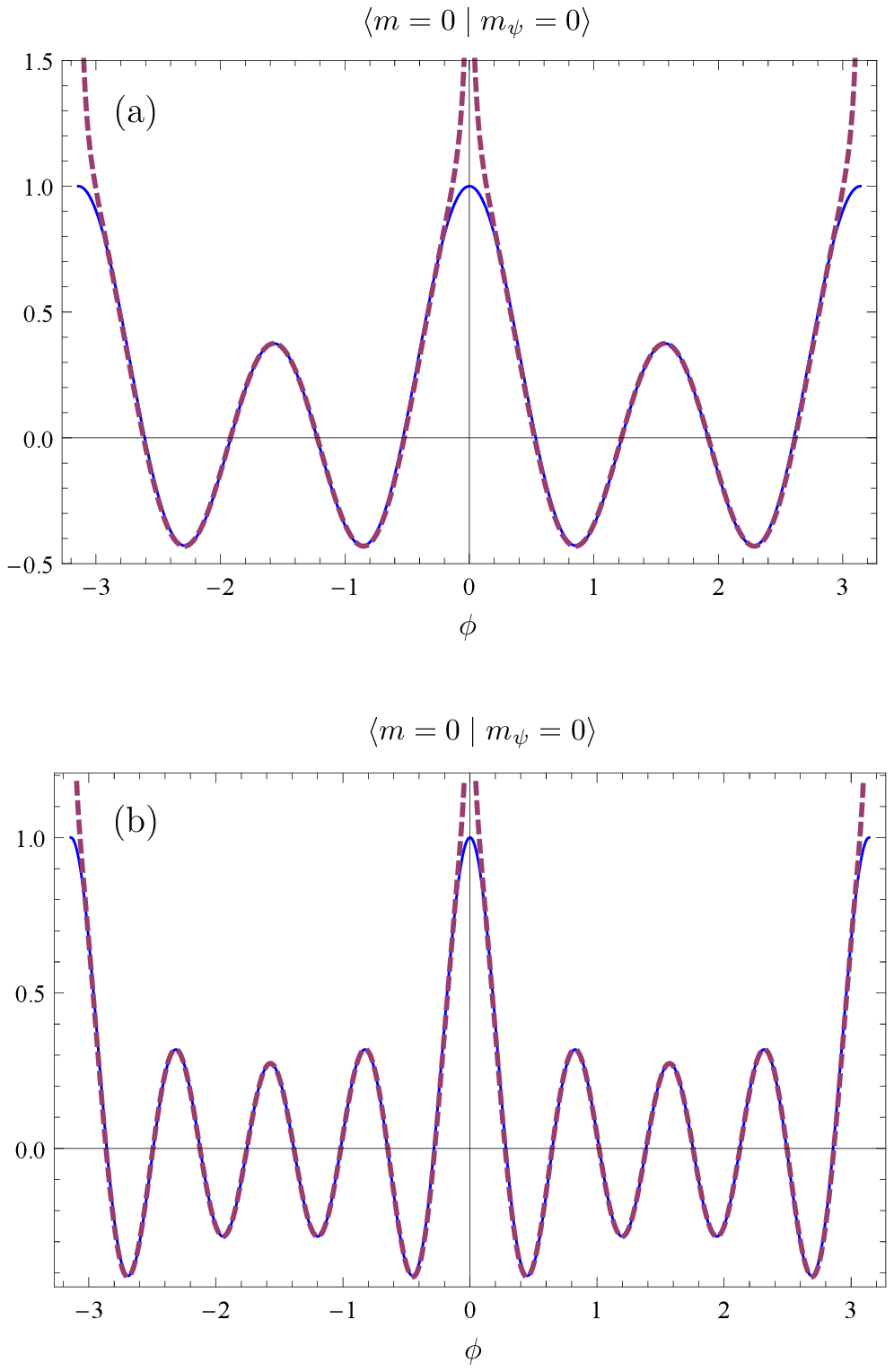}}}}
\end{picture}
\vspace{-2.5cm}
\caption{\label{pairs}
Comparison of exact solution and approximate separation of the action $S$ and the envelope function $A$ for (a) eight photons and (b) sixteen photons. The dotted lines show the results of the approximation, while the full line is the precise result obtained from the complete state vector in the nine and seventeen dimensional Hilbert spaces, respectively. The approximate amplitude $\langle m \mid m_\psi \rangle$ only diverges from the exact solution close to $\phi=0$ and $\phi=\pi$, where the quantum mechanical solution is limited to a probability of one, while the classical amplitude describes a probability density that diverges to infinity.
}
\end{figure}

We can now find an approximate expression for the amplitude of the $\langle m \mid \psi \rangle$ in the case of equal photon numbers in the input and output. There is only the problem of the integration constant for the action $S$. We can solve this problem by noting that $\phi=\pi/2$ corresponds to the action of a 50:50 beam splitter, where it is known that the probabilities of finding an odd number of photons in the output ports is zero for equal photon numbers in the input. This means that, at $\phi=\pi/2$, $\cos(S/\hbar)=0$ for odd $N/2$ and $\cos(S/\hbar)=\pm 1$ for even $N/2$. The phase dependence of $S$ is therefore given by
\begin{equation}
S(0,0,\phi) = - \frac{\hbar}{2}\left((N+1) \phi - \frac{\pi}{2}\right).
\end{equation}
Note that this solution is obtained for phases between zero and $\pi$, since the approximation that $S$ changes more rapidly than $A$ breaks down near $\phi=0$ and near $\phi=\pi$. To include the range between $\phi=-\pi$ and $\phi=0$, it is convenient to use the absolute value of $\phi$, for an approximate solution of 
\begin{equation}
\langle m=0 \mid \psi(m_\psi=0) \rangle \approx
2 \sqrt{\frac{1}{\pi (N+1) \sin(|\phi|)}} \cos\left(\frac{N+1}{2}|\phi| - \frac{\pi}{4}\right). 
\end{equation}
Fig. \ref{pairs} shows the comparison between the exact solution and the approximation for eight photons and for sixteen photons. As the figure shows, the approximation correctly describes the amplitudes $\langle m \mid \psi \rangle$ except in the immediate vicinity of $\phi=0$ and $\phi=\pm \pi$, where the approximation diverges to values higher than one. This discrepancy is easy to understand, since the classical approximation describes a probability density instead of a discrete probability, so it can exceed a value of one if the probability distribution in $J_\phi$ is narrower than $\hbar$. Significantly, the discreteness of the outcomes has no effect on the results outside of these two very narrow regions. Even for eight photons, the effects of higher order terms in $\hbar$ is mostly negligible.

\begin{figure}[th]
\vspace{-1.5cm}
\begin{picture}(500,480)
\put(0,0){\makebox(480,480){
\scalebox{0.8}[0.8]{
\includegraphics{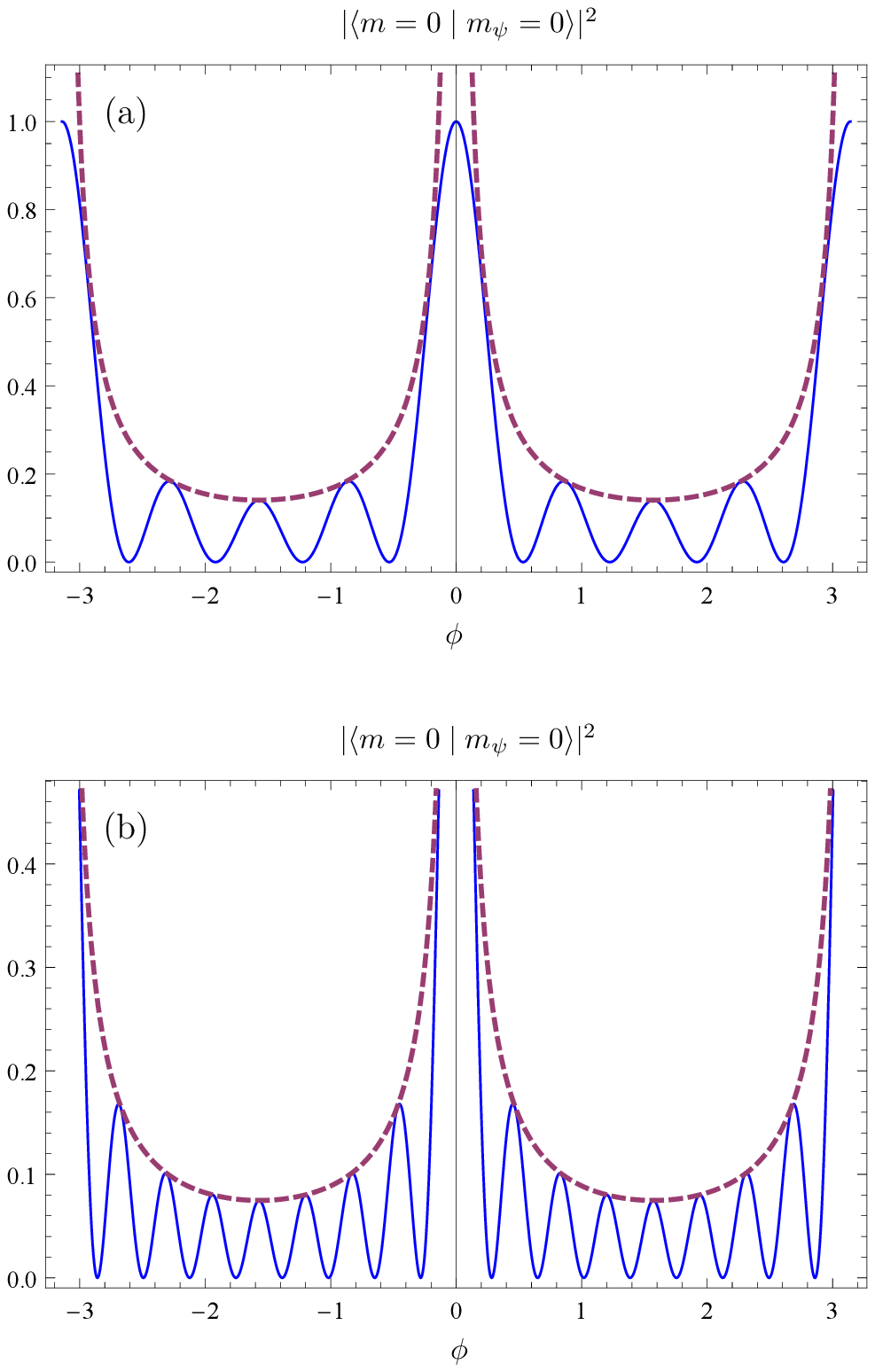}}}}
\end{picture}
\vspace{-2.5cm}
\caption{\label{rates}
Experimentally observable multi-photon interference fringes for equal photon numbers in the input and the output. (a) shows the phase dependent count rates for eight photons and (b) shows the phase dependent count rate for sixteen photons. The dotted line shows the envelope function $4 A^2$ to indicate that the amplitude of the fringes follows the classical statistics of random phase interference. 
}
\end{figure}

Since the experimental count rates of a multi-photon interference experiment are given by the probabilities $|~\langle~m~\mid~\psi~\rangle~|^2$, these are shown in Fig. \ref{rates}. Note that the envelope values indicated in the figure correspond to twice the classical probability densities expected for random phase interference with the same input and output conditions. The coincidence of this envelop with the peaks of the interference fringes in Fig. \ref{rates} thus illustrates how classical statistics emerge when the phase resolution fails to resolve the quantum interferences associated with the action $S$. As mentioned above experimental data of this type has already been reported, e.g. in \cite{Xia13} for a six photon state, where the minima are separated by phases of $\Delta \phi = 2 \pi/7$. In general, the positions of the minima are found at $\cos(S)=0$, which means that the minima for $0<\phi<\pi$ are located at
\begin{equation}
\phi_{\mbox{min.}}=\frac{3 \pi}{2 (N+1)},  \frac{7 \pi}{2 (N+1)}, \frac{11 \pi}{2 (N+1)}, \ldots
\end{equation}
For $-\pi < \phi < \pi$, there will be a total of $N$ minima, equal to the total number of photons. However, the width of the peaks at $\phi=0$ and at $\phi=\pm \pi$ is 1.5 times the width of the other peaks, resulting in $N-2$ interference fringes with a width of $\Delta \phi=2 \pi/(N+1)$ and two fringes with a width of $\Delta \phi^\prime=3 \pi/(N+1)$.

\begin{figure}[th]
\vspace{-1cm}
\begin{picture}(500,240)
\put(0,0){\makebox(480,240){
\scalebox{0.8}[0.8]{
\includegraphics{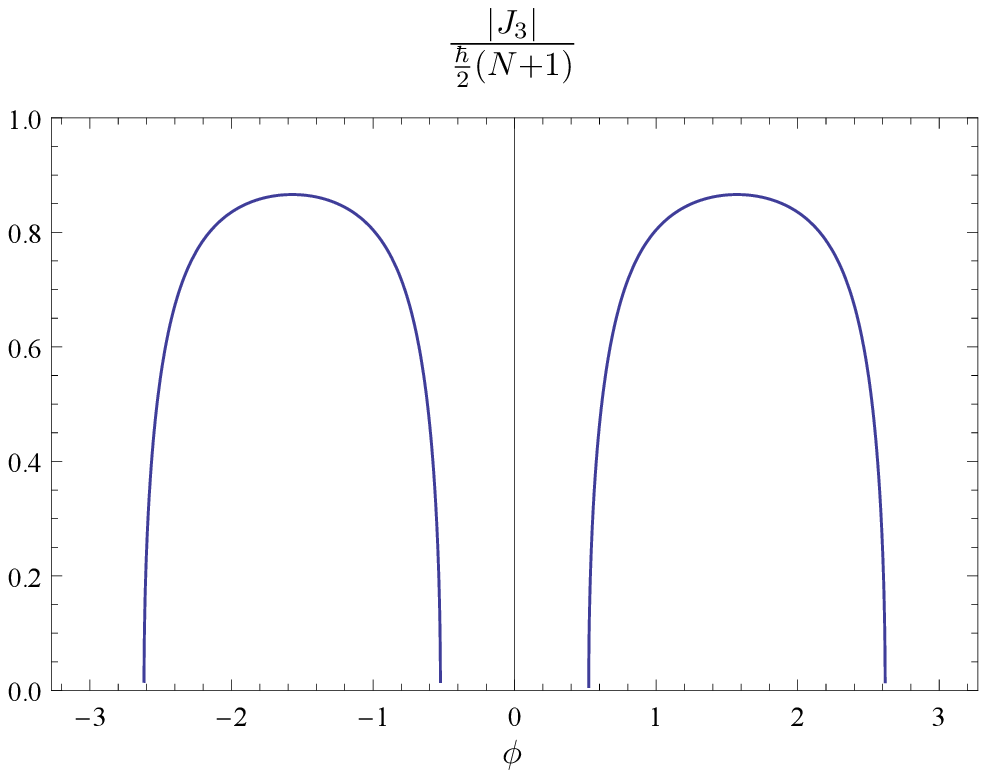}}}}
\end{picture}
\vspace{-1.2cm}
\caption{\label{J3cross}
Dependence of the intensity difference between the paths in the interferometer $|J_3|$ on phase $\phi$ for an input photon number difference of $2m_\psi=0$ and an output photon number difference of $2m = N/2$. The value is given relative to the total length of the $J$-vector, which is $\hbar (N+1)/2$.
}
\end{figure}

\begin{figure}[th]
\vspace{-1.5cm}
\begin{picture}(500,480)
\put(0,0){\makebox(480,480){
\scalebox{0.8}[0.8]{
\includegraphics{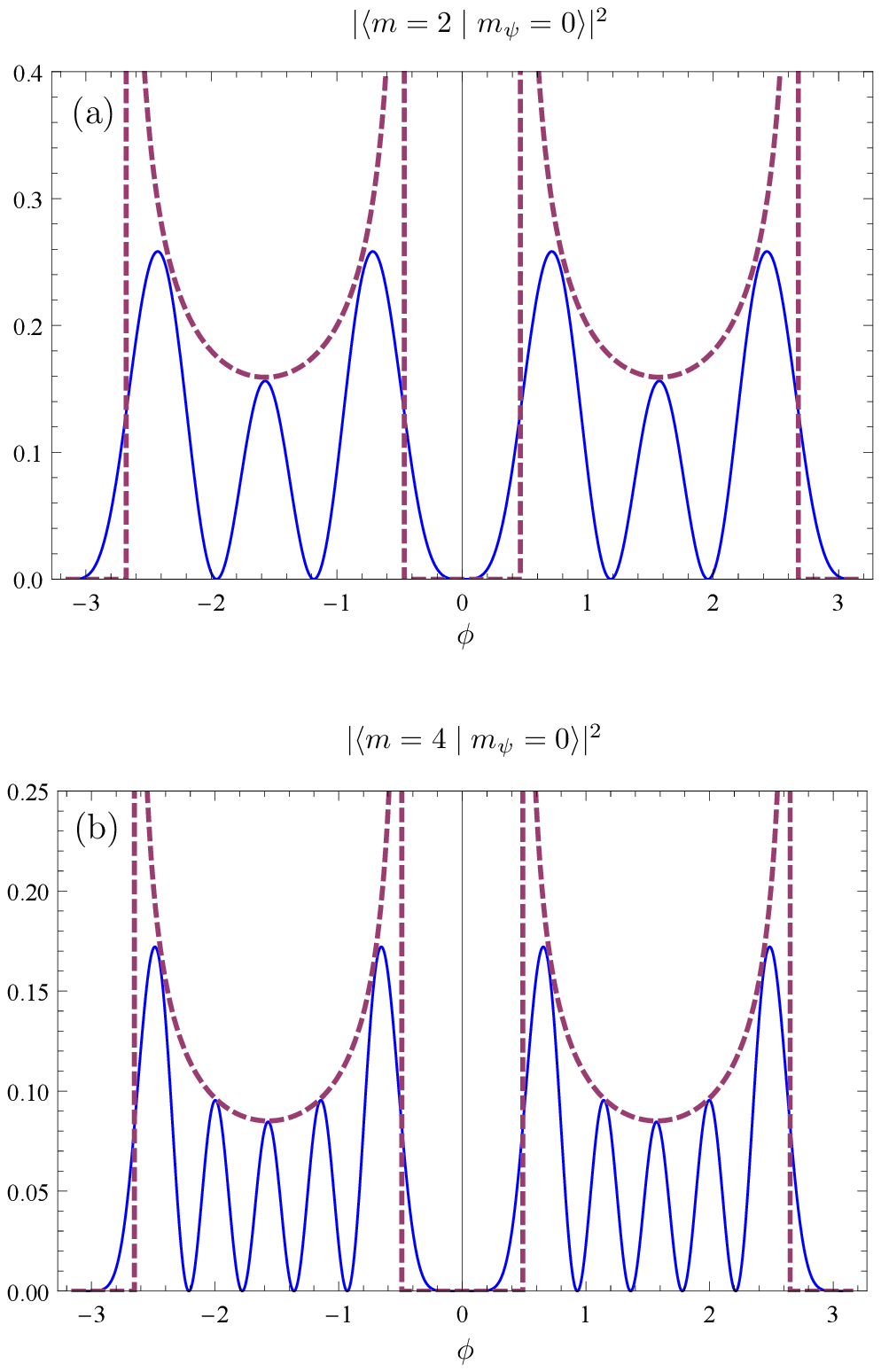}}}}
\end{picture}
\vspace{-2.5cm}
\caption{\label{Pcross}
Experimentally observable multi-photon interference fringes for equal photon numbers in the input and a photon number difference of $2m=N/2$ in the output. (a) shows the phase dependent count rates for eight photons and (b) shows the phase dependent count rate for sixteen photons. The dotted line shows the envelope function $4 A^2$ to indicate that the amplitude of the fringes follows the classical statistics of random phase interference. 
}
\end{figure}

Next we consider cases where the photon number difference is non-zero in either the input or the output. Experimentally, it is relatively easy to obtain this kind of data from an $m_\psi=0$ input state by simply arranging the detectors to detect a non-zero photon number difference of $2 m$ between the output ports. From the output photon number difference $2 m$, we can then estimate the intensity difference between the paths in the interferometer as
\begin{equation}
\label{eq:m0}
|J_3(\hbar m,0,\phi)| = \hbar \sqrt{\frac{N(N+2)}{4} - \frac{m^2}{(\sin(\phi))^2}}
\end{equation}
Since a minimal phase shift of about $\phi = \arcsin(m/(N+1))$ is necessary to achieve an output photon number difference of $m$ by optical interference, this equation only has valid solutions for a limited range of phase shifts centered around $\phi=\pi/2$. In all cases, the magnitude of $m$ limits the maximal value of $J_3$, and hence the periodicity of the multi-photon interference fringes. Fig. \ref{J3cross} shows the dependence of $|J_3|$ on phase for $m \approx (N+1)/4$. In this case, the maximal value at $\phi=\pi/2$ is equal to $\sqrt{3/4}$ times the maximal value of $\hbar (N+1)/2$.

Although it is possible to find an analytical solution for the action, the merit of the approximation is that it provides an insight into the physics that determines the periodicity of multi-photon interferences, and into the relation between the action and the envelope function that describes the classical statistics of random phase interference. We will therefore analyze both aspects using the exact solutions of the interference fringes obtained for specific photon numbers. Fig. \ref{Pcross} (a) shows the interference fringes for eight photons and $m=N/4=2$ in the output, and Fig. \ref{Pcross} (b) shows the fringes for sixteen photons and $m=N/4=4$ in the output. The approximate envelop function given by Eq.(\ref{eq:ergo}) is also indicated, showing that the approximation works well within the range of classical solutions. Outside this range, there is a non-zero evanescent solution that quickly drops to zero. Within the range of classical solutions, we observe multi-photon interference effects in the form of fringes of width $\Delta \phi_{\mathrm{exp.}}$. This width is easy to observe experimentally, since it can be defined as the phase difference between two consecutive minima of the output count rate. We can use this definition to obtain an experimental value of $|J_3|$ for a specific fringe,
\begin{equation}
\label{eq:evaluate}
|J_3|_{\mathrm{exp.}} = \frac{\pi}{\Delta \phi_{\mathrm{exp.}}}.
\end{equation} 
We now apply this method of evaluation to the exact solutions of $|\langle m \mid \psi \rangle|^2$ in order to find out how well the approximation given by Eq.(\ref{eq:J3}) corresponds to the actual periodicities of multi-photon interference. For the eight photon case shown in Fig. \ref{Pcross} (a), probabilities of zero occur at $\phi=1.183$ and at $\phi=1.958$, resulting in a single interference fringe with $\Delta \phi_{\mathrm{exp.}}=0.775$. The corresponding value of $|J_3|$ is $4.05 \hbar$. According to Eq. (\ref{eq:m0}), the maximal possible value of $|J_3|$ for $N=8$ and $m=2$ should be $4.00$, indicating that the approximation slightly underestimates the value of $|J_3|$.

For the sixteen photon case shown in Fig. \ref{Pcross} (b), probabilities of zero occur at $\phi=0.931$, $\phi=1.362$, $\phi=1.780$, and $\phi=2.211$, resulting in three interference fringes with $\Delta \phi_{\mathrm{exp.}}=0.431$, $\Delta \phi_{\mathrm{exp.}}=0.418$, and $\Delta \phi_{\mathrm{exp.}}=0.431$. The corresponding values of $|J_3|$ are $7.29 \hbar$, $7.52 \hbar$, and $7.29 \hbar$. As expected from the theory, $|J_3|$ is maximal around $\phi=\pi/2$, although the theory again underestimates the value, with $|J_3|=7.48 \hbar$ at $\phi=\pi/2$ for $N=16$ and $m=4$. As shown in Fig.\ref{J3cross}, the value of $|J_3|$ drops off symmetrically as the phase is shifted from $\phi=\pi/2$ to higher or lower phases. We can use Eq.(\ref{eq:m0}) to find the phase at which the expected value of $|J_3|$ is $7.29 \hbar$. We obtain phases of $\phi=1.263$ and $\phi=1.879$. Both of these phases are located near the center of the interference fringes from which the $|J_3|$ values are derived.

In summary, the comparison between the approximate separation of quantum interference and (classical) envelope function can also be confirmed for different photon numbers in either the input or the output. Specifically, we can explain both the confinement of output probabilities to a finite phase interval around $\phi=\pi/2$ and the width of quantum interference fringes using the input-output relations given by Eqs.(\ref{eq:J3}) and Eq.(\ref{eq:ergo}). Remarkably, both of these relations also apply to classical random phase interference, although classical statistics would result in an output probability given by $2 A^2$, without the quantum interference effects associated with the value of $|J_3|$.

\begin{figure}[th]
\vspace{-1cm}
\begin{picture}(500,240)
\put(0,0){\makebox(480,240){
\scalebox{0.8}[0.8]{
\includegraphics{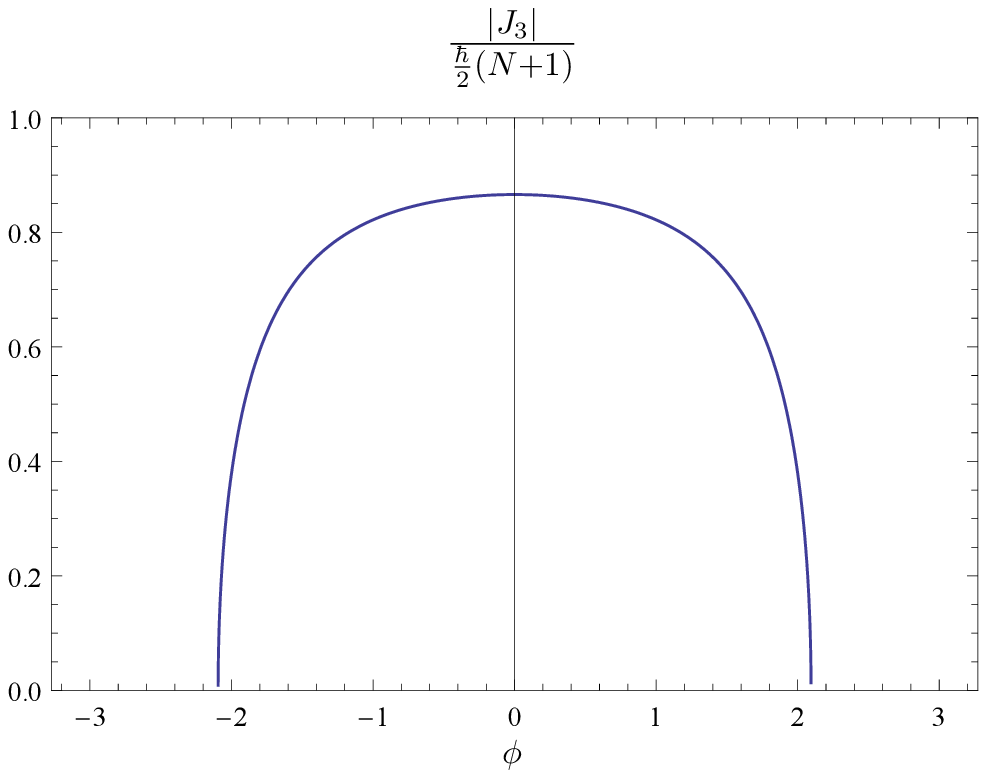}}}}
\end{picture}
\vspace{-1.2cm}
\caption{\label{J3self}
Dependence of the intensity difference between the paths in the interferometer $|J_3|$ on phase $\phi$ for an input photon number difference of $2m_\psi=N/2$ and an output photon number difference of $2m = N/2$. The value is given relative to the total length of the $J$-vector, which is $\hbar (N+1)/2$.
}
\end{figure}

\begin{figure}[th]
\vspace{-1.5cm}
\begin{picture}(500,480)
\put(0,0){\makebox(480,480){
\scalebox{0.8}[0.8]{
\includegraphics{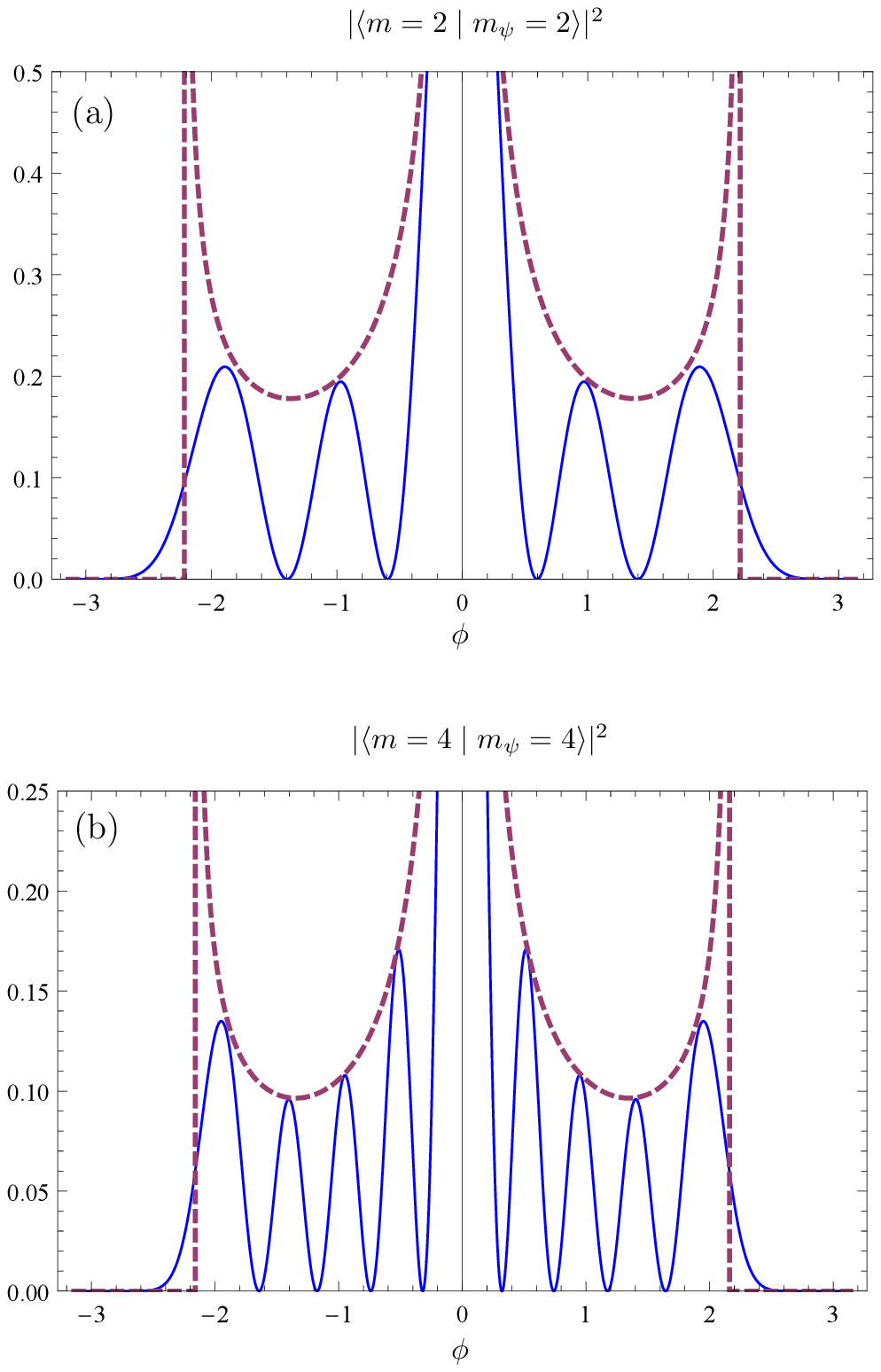}}}}
\end{picture}
\vspace{-2.5cm}
\caption{\label{Pself}
Experimentally observable multi-photon interference fringes for photon number differences of $2m_\psi=2m=N/2$ in both the input and the output. (a) shows the phase dependent count rates for eight photons and (b) shows the phase dependent count rate for sixteen photons. The dotted line shows the envelope function $4 A^2$ to indicate that the amplitude of the fringes follows the classical statistics of random phase interference. 
}
\end{figure}

Finally, it may also be good to consider the case where both the input and the output have a non-vanishing photon number difference between the ports. To keep the mathematics simple, we choose the case of $m=m_\psi$, where the output photon number distribution happens to be identical to the input photon number distribution. The intensity difference $|J_3|$ between the paths is then given by  
\begin{equation}
\label{eq:mm}
|J_3(\hbar m,m,\phi)| = \hbar \sqrt{\frac{N(N+2)}{4} - \frac{2 m^2}{1+\cos(\phi)}}.
\end{equation}
Note that the maximal value of $|J_3|$ is the same as the maximal value in Eq.(\ref{eq:m0}), but it occurs at $\phi=0$ instead of at $\phi=\pi/2$. The value of $|J_3|$ drops continuously as $|\phi|$ increases, until it reaches zero at $1+\cos(\phi) \approx 8 m^2/(N+1)^2$. Fig. \ref{J3self} shows the dependence of $|J_3|$ on phase for $m\approx(N+1)/4$, where the value of $|J_3|$ reaches zero at $|\phi|=2 \pi/3$.

Fig. \ref{Pself} (a) shows the interference fringes for eight photons and $m=N/4=2$ in both input and output, and \ref{Pself} (b) shows the fringes for sixteen photons and $m=N/4=4$ in input and output. The approximate envelope function is also shown, indicating that Eq.(\ref{eq:ergo}) is a good approximation of the precise results up to the limit of validity near $|\phi|=2 \pi/3$. We can now evaluate the width of the fringes and obtain the corresponding values of $|J_3|$ according to Eq.(\ref{eq:evaluate}). For the eight photon case shown in Fig. \ref{Pself} (a), probabilities of zero occur at $\phi=0.597$ and at $\phi=1.397$, resulting in a single interference fringe with $\Delta \phi_{\mathrm{exp.}}=0.800$. The corresponding value of $|J_3|$ is $3.93\hbar$. According to Eq.(\ref{eq:mm}), this value of $|J_3|$ is obtained at a phase of $|\phi|=0.713$. Although this phase is found within the fringe, it is much closer to the low phase minimum than to the high phase minimum, indicating that the approximation slightly underestimates the $|J_3|$ value. In the middle of the fringe, at $\phi=0.997$, Eq.(\ref{eq:mm}) assigns a $|J_3|$ value of only $3.85 \hbar$. This is consistent with the result from Fig.(\ref{Pcross}) (a) above, suggesting an error margin of about two percent for eight photon fringe widths obtained from Eq.(\ref{eq:J3}). 

For the sixteen photon case shown in Fig. \ref{Pself} (b), probabilities of zero occur at $\phi=0.321$, $\phi=0.740$, $\phi=1.175$, and $\phi=1.644$, resulting in three interference fringes with $\Delta \phi_{\mathrm{exp.}}=0.419$, $\Delta \phi_{\mathrm{exp.}}=0.435$, and $\Delta \phi_{\mathrm{exp.}}=0.469$. The corresponding values of $|J_3|$ are $7.50\hbar$, $7.22\hbar$, and $6.70\hbar$. As expected, the values of $|J_3|$ decrease as the phase shift increases. The $|J_3|$ value of the first fringe exceeds the maximal value of $|J_3|=7.48 \hbar$ obtained from Eq.(\ref{eq:mm}), indicating once more that the theory underestimates the values of $|J_3|$. For the other two fringes, the values of $|J_3|$ obtained for their respective widths are found at $|\phi|=0.914$ for $|J_3|=7.22\hbar$, and at $|\phi|=1.389$ for $|J_3|=6.70\hbar$. Both results are very close to the centers of the fringes, confirming that Eq.(\ref{eq:HJE}) provides a good approximate description of multi-photon interferences.

The results obtained for all examples show that the physics of multi-photon interference can be understood in terms of the separation between the action $S$ that describes quantum interferences and the amplitude $A$ that describes the classical statistics of random phase interference. In particular, it is possible to explain the periodicity of multi-photon interference in terms of the intensity difference between the paths expected from the {\it combination} of input and output conditions. It is therefore important to recognize that multi-photon interference is not just the result of superpositions in the input state, but also involves quantum coherences in the multi-photon statistics of the measurement outcome. In the case of photon number states in the input, the relation is completely symmetric, and the same phase dependent interference fringes will be obtained when input and output are exchanged.

\section{The role of the action in multi-photon interference}
\label{sec:action}

In the analysis presented above, we have studied the multi-photon interference fringes of photon number state inputs in the two input ports of a two-path interferometer. However, the approach developed at the start of Sec. \ref{sec:wv} is not limited to these specific states. Specifically, Eqs.(\ref{eq:action}) and (\ref{eq:quadact}) are generally valid expressions of the phase evolution of interference fringes in terms of the weak values of $\hat{J}_3$ and $\hat{J}_3^2$. The separation of amplitude $A$ and action $S$ introduced in Eq.(\ref{eq:separate}) is possible whenever the amplitudes have only real values. This means that the real part of the weak values of $\hat{J}_3$ must be zero for all phases $\phi$,
\begin{equation}
\label{eq:cond1}
\mbox{Re}\left(\frac{\langle m \mid \hat{J}_3 \mid \psi \rangle}{\langle m \mid \psi \rangle}\right)=0.
\end{equation}
Since the phases of different eigenstate components of $\hat{J}_3$ evolve differently, this criterion cannot be satisfied unless each pair of components with opposite eigenvalues have the same amplitude and opposite phases. If the eigenstates of $\hat{J}_3$ with eigenvalues of $\hbar m_3$ are written as $\mid m_3 \rangle$, this criterion can be expressed by the condition
\begin{equation}
\langle m \mid m_3 \rangle \langle m_3 \mid \psi \rangle = \langle \psi \mid -m_3 \rangle \langle -m_3 \mid m \rangle. 
\end{equation}
Usually, this condition is satisfied separately by both the input state and the output state, so that both states are symmetric in the path basis. Specifically, the eigenstates of $\hat{J}_\phi$ that represent the results of photon number measurements in the output ports all satisfy the  condition
\begin{equation}
\langle m \mid m_3 \rangle = \langle -m_3 \mid m \rangle.
\end{equation}
Therefore, the amplitudes $\langle m \mid \psi \rangle$ will be real at all phases $\phi$ if the $\hat{J}_3$ components of $\mid \psi \rangle$ satisfy
\begin{equation}
\label{eq:pathsym}
\langle m_3 \mid \psi \rangle = \langle \psi \mid -m_3 \rangle. 
\end{equation}
Note that it has been shown elsewhere that input states satisfying this path symmetry condition achieve their maximal phase sensitivity in photon number measurements \cite{Hof09}. The present analysis investigates the more detailed mechanism by which this is achieved. 

In general, the approximate solution of Eq.(\ref{eq:quadact}) is given by Eq.(\ref{eq:second}), so that the multi-photon interference fringes can be described by the phase dependent action $S$ with
\begin{equation}
\label{eq:act2}
\frac{\partial}{\partial \phi} S(\phi) = -\sqrt{\frac{\langle m \mid \hat{J}_3^2 \mid \psi \rangle}{\langle m \mid \psi \rangle}}
\end{equation}
It is therefore possible to derive the periodicity of the interference fringes from the weak value of the squared phase shift generator $\hat{J}_3^2$ for any path symmetric state. Note that Eq.(\ref{eq:quadact}) indicates that the weak value of $\hat{J}_3^2$ is always real if the states and measurements satisfy Eq.(\ref{eq:cond1}) at all phases. In the limit of high photon number $N$, it is usually possible to find a classical relation that defines the value of $J_3$ as a function of initial conditions $\psi$, measurement outcome $m$, and phase shift $\phi$. It is then possible to relate the multi-photon interference effects to the deterministic relations between initial and final conditions known from classical wave interference. 

In the present case, the action $S$ is directly observable as an interference fringe because the amplitudes $\langle m \mid \psi \rangle$ are all real. This is a result of the symmetry between negative and positive intensity differences between the paths. If there is only a single classical solution for $J_3$, the action given by Eq.(\ref{eq:HJE}) will merely describe an unobservable phase evolution of $\langle m \mid \psi \rangle$. In this sense, one can explain quantum interference as an interference between the two classical solutions $+J_3$ and $-J_3$, both of which connect $\psi$ to $m$ under phase shifts of $\phi$. Importantly, this interference can be described without any reference to Hilbert space vectors, since the physical properties $J_i$ can all be defined within classical wave theory. It may therefore be useful to take a closer look at the relation between classical wave interference and the quantum interference of multi-photon states.

\section{Classical interference versus quantum interference}
\label{sec:NOON}

As shown above, multi-photon interference can be understood in terms of the weak values of the generator $\hat{J}_3$ that describes the phase shift induced by a two-path interferometer. To explain the connection with more conventional ideas of quantum interference as an effect of superpositions in the initial state, it may be useful to consider a particularly simple solution of Eq.(\ref{eq:act2}), where the periodicity of the interference fringes is determined by the input state and does not depend on output photon number or phase. This situation only occurs if the input state is an eigenstate of $\hat{J}_3^2$, so that the weak value is given by the initial eigenvalue and does not depend on phase anymore. Except for arbitrary phase shifts, the complete set of states with constant fringe periodicity is therefore given by
\begin{equation}
\label{eq:super}
\mid \psi(m_3) \rangle = \frac{1}{\sqrt{2}} \left(\mid m_3 \rangle + \mid - m_3 \rangle\right), 
\end{equation}
where $\mid m_3 \rangle$ are the $\hat{J}_3$ eigenstates with eigenvalues of $\hbar m_3$. Similar to the case of equal photon numbers in input and output, the derivative of the action is constant and a multiple of $\hbar$,
\begin{equation}
\frac{\partial}{\partial \phi} S(\phi) = - \hbar m_3.
\end{equation}
For an $N$-photon state, the maximal number of fringes is obtained for the so-called NOON state, where $m_3=N/2$. It is therefore tempting to think of multi-photon interference as an $N$-fold increase in the number of fringes, or an $N$-fold enhancement of phase sensitivity. However, even the simple superposition states given by Eq.(\ref{eq:super}) show that the number and the periodicity of fringes is not a function of photon number, but a function of photon number distribution between the paths. In general, the weak value of $\hat{J}_3^2$ provides an estimate of the intensity distribution between the paths for a specific combination of input state, measurement outcome, and phase shift. This estimate identifies the proper physics of multi-photon interference. 

What is perhaps most remarkable about the results presented above is that the quantum effects of multi-photon interference can be explained by using the deterministic relations between field coherences already known from classical two-path interference. The classical interference effect is actually described by the operator relation in Eq.(\ref{eq:Heis}), where the output intensities are expressed as a function of input intensities, coherence, and phase shift. In classical theory, probabilities enter the picture only if there is some randomness in the initial conditions. In the case of photon number inputs discussed above, this randomness can be identified as a random phase difference between the two input fields. As the approximation shows, the quantum mechanical result reproduces this classical probability distribution over random phases in the form of the envelope function that describes the amplitude of the quantum interference fringes. 

Importantly, quantum interference is not an effect that emerges gradually from classical field interference as photon number increases, but should be considered as a fundamentally different effect associated with the non-classical relation between the phase shift $\phi$ and its generator $\hat{J}_3$. The only reason why this effect is related to photon number is that, as a quantum effect, its scale is determined by $\hbar$. All classical effects scale with intensity, only the quantum interference fringes decrease in width as photon number increases. The analysis in this paper shows that it is possible to separate the quantum scale and the classical scale in the quantum theory of multi-photon interference fringes, highlighting the different physics involved in the two distinct processes. 

\section{Conclusions}
\label{sec:concl}

In conclusion, multi-photon interferences can be explained in terms of the weak values of the intensity difference between the paths of the interferometer. For photon number states, it is then possible to describe each multi-photon interference fringe by its own differential equation. The approximate solution of this differential equation identifies the periodicity of the fringes with the intensity difference between the paths inside the interferometer, which can be determined from the classical relation between input intensity difference, output intensity difference, and phase. 

We have shown that it is possible to separate the quantum effects of multi-photon interference from the classical statistics of random phase interference. Specifically, we can trace quantum interference back to the action of the phase shift, which is given by the product of the generator $\hat{J}_3$ and the phase shift $\phi$. The photon number appears in this relation only because $\hat{J}_3$ is quantized in units of $\hbar$. The classical units of intensity difference are action units, where $\hbar$ is the ratio between the action and the quantum phase. Our analysis thus shows that an detailed analysis of multi-photon interference can be achieved by focusing on the classical field properties if both the input state and the measurement outcomes are treated on an equal footing. 

Our work provides a general characterization of multi-photon interference that may serve as a road map to future experiments involving larger photon numbers. We have identified the relevant features of different two-path interference scenarios and explained the origin and the physical meaning of the experimentally observable fringe width. We believe that these observations will be extremely helpful in the characterization of multi-photon quantum devices and in the development of new approaches to multi-particle quantum physics. 

\section*{Acknowledgment}
This work was supported by
JSPS KAKENHI Grant Number 26220712.


\end{document}